\documentclass{atlasnote}
\usepackage{anysize}
\marginsize{3cm}{2cm}{2cm}{2cm}
\usepackage{amsmath}
\usepackage{mathtools}
\usepackage{subfigure}
\usepackage{graphicx}

\title{Asymptotic Formula for a General Double-Bounded Custom-Sided Likelihood Based Test Statistic}

\usepackage{authblk}

\author{Will Buttinger}

\begin{document}
\begin{center}%
    {\Large\bf Asymptotic Formula for a General Double-Bounded Custom-Sided Likelihood Based Test Statistic \par}%
    \vskip 3em%
    {\large
     \lineskip .75em%
      \begin{tabular}[t]{c}%
        Will Buttinger\\
        {\small University of Cambridge}
      \end{tabular}\par}%
  \end{center}\par
  \begin{center}
    {\bfseries \abstractname}
    \quotation
    This paper presents the asymptotic distributions of a general likelihood-based test statistic, derived using results of Wilks and Wald. The general form of the test statistic incorporates the test statistics and associated asymptotic formulae previously derived by Cowan, Cranmer, Gross and Vitells, which are seen to be special cases of the likelihood-based test statistic described here. 
    \endquotation
  \end{center}




\section{Introduction}
This paper defines likelihood based test statistics for hypothesis tests of a particular parameter of interest, where the parameter is believed to be physically (or otherwise) constrained to lie within a certain range. Defining both an upper and lower boundary for the parameter of interest will be referred to as a double-bounded test statistic, and (in the interest in defining a convention) will be notated by a letter with a tilde and a bar above it, e.g. $\bar{\tilde{t}}_\mu$ (the $\mu$ labels the value of the parameter of interest that is to be tested with this particular instance of the test statistic). In the special case where the upper boundary is moved to $\infty$, the resulting test statistic is referred to as lower-(single)bounded, notated by a letter with just a tilde above it (e.g. $\tilde{t}_\mu$). For the case where the lower boundary is moved to $-\infty$, the notation $\bar{t}_\mu$ will be used - this is an upper-(single)bounded test statistic. 

Likelihood-based test statistics can also be either two-sided or one-sided. This nomenclature refers to whether data resulting in a maximum likelihood estimator for the parameter of interest ($\hat\mu$) that is not equal to the value being tested ($\mu$) should be considered as increasingly incompatible with the hypothesis for all values of $\hat\mu$ (a two-sided test statistic), or instead if $\hat\mu$ in only  one-direction relative to $\mu$ should be considered increasingly incompatible (a one-sided test statistic). The convention that will be used to distinguish these two cases is that two-sided test statistics use the letter $t$, e.g. $\bar{\tilde{t}}_\mu$, whereas one-sided test statistics will use the letter $q$, e.g. $\bar{\tilde{q}}_\mu$. 

It is possible to define any customized region of $\hat\mu$ values that are considered fully compatible with a given value of the parameter of interest $\mu$. This type of test statistic will be referred to as a custom-sided test statistic, and the convention will be to use the letter $k$ for such statistics. The definition of the custom-sided test statistic should therefore identify what regions of $\hat{\mu}$ values are compatible with the given value of $\mu$ being tested (and therefore, where the test statistic should become 0, corresponding to maximal compatibility for likelihood ratio based test statistics). These regions of compatibility will be denoted by $m_\mu(\hat{\mu})$, a function that is valued 1 for $\hat{\mu}$ considered compatible with $\mu$, and 0 otherwise. Therefore a two-sided test statistic can be expressed as a custom-sided test statistic with
\begin{equation}
  m_\mu(\hat{\mu}) = 0 \; \forall \hat{\mu} 
\end{equation}
A one-sided test statistic is a custom-sided test statistic with 
\begin{equation}
  m_\mu(\hat{\mu}) = \begin{cases} 
                      0 & \hat{\mu}\leq \mu \\
                      1 & \hat{\mu}>\mu.
                     \end{cases}
\end{equation}
Asymptotic distributions for lower bound (at $\mu=0$) one-sided, two-sided and double-bounded two-sided likelihood-based test statistics have been previously derived by Cowan, Cranmer, Gross and Vitells (\cite{asymptoticpaper} and \cite{asymptoticpaper2}). This paper presents the asymptotic distribution of a more general likelihood-based test statistic $\bar{\tilde{k}}_\mu$, a double-bounded custom-sided test statistic. The distribution is derived in section~\ref{sec:derivation} using approximate methods based on results due to Wilks~\cite{wilkspaper} and Wald~\cite{waldpaper}. Section~\ref{sec:dist} summarises the test statistic and the resulting asymptotic cumulative probability distribution and probability density function, including how previously derived asymptotic distributions can be recovered as special cases of the $\bar{\tilde{k}}_\mu$ test statistic. Section~\ref{sec:example} presents a Monte Carlo study for a hypothetical analysis, designed to test the validity of the asymptotic formula that is derived in section~\ref{sec:dist}.

\section{Distribution of $\bar{\tilde{k}}_\mu$}
\label{sec:dist}

The double-bounded custom-sided test statistic $\bar{\tilde{k}}_\mu$ is defined as
\begin{equation}
 \bar{\tilde{k}}_\mu = \begin{cases} 
  0 & m_\mu(\hat{\mu})=1, \\
  -2\ln{\frac{L(\mu,\hat{\hat{\theta}}(\mu))}{L(\mu_L,\hat{\hat{\theta}}(\mu_L))}} & m_\mu(\hat{\mu})=0,\hat{\mu}<\mu_L,\\
  -2\ln{\frac{L(\mu,\hat{\hat{\theta}}(\mu))}{L(\hat{\mu},\hat{\theta})}} & m_\mu(\hat{\mu})=0,\mu_L<\hat{\mu}<\mu_H,\\
  -2\ln{\frac{L(\mu,\hat{\hat{\theta}}(\mu))}{L(\mu_H,\hat{\hat{\theta}}(\mu_H))}} & m_\mu(\hat{\mu})=0,\hat{\mu}>\mu_H,
\end{cases}
\label{eq:teststatdef}
\end{equation}
where the \emph{compatibility function} $m_\mu(\hat\mu)$ is defined as:
\begin{equation}
 m_\mu(\hat\mu) = \begin{cases} 
                   1 & \hat\mu\,\text{ is considered compatible with }\,\mu,\\
                   0 & \text{otherwise}.
                  \end{cases}
\end{equation}
The asymptotic cumulative distribution of $\bar{\tilde{k}}_\mu$, under the Wald approximation (where $\hat\mu$ follows a Gaussian distribution with a mean $\mu'$ and standard deviation $\sigma$) is found to be (see section~\ref{sec:derivation}):
\begin{eqnarray}
 F(\bar{\tilde{k}}_\mu|\mu') & = &
\left(\Phi\left(\sqrt{\bar{\tilde{k}}_\mu}-\Lambda_y\right)+\Phi_m\left(\Lambda_y-\sqrt{\bar{\tilde{k}}_\mu}\right)\right)\left(1-H(\bar{\tilde{k}}_\mu-\bar{\tilde{k}}_\mu^L)\right) \nonumber \\
&  & +
\left(\Phi\left(\sqrt{\bar{\tilde{k}}_\mu}+\Lambda_y\right)-\Phi_m\left(\Lambda_y+\sqrt{\bar{\tilde{k}}_\mu}\right)\right)\left(1-H(\bar{\tilde{k}}_\mu-\bar{\tilde{k}}_\mu^H)\right) \nonumber \\
&  & +
\left(\Phi\left(\frac{\bar{\tilde{k}}_\mu-\Lambda_L}{\sigma_L}\right)+\Phi_m\left(\frac{\Lambda_L-\bar{\tilde{k}}_\mu}{\sigma_L}\right)\right)H(\bar{\tilde{k}}_\mu-\bar{\tilde{k}}_\mu^L) \nonumber \\
&  & +
\left(\Phi\left(\frac{\bar{\tilde{k}}_\mu-\Lambda_H}{\sigma_H}\right)-\Phi_m\left(\frac{\bar{\tilde{k}}_\mu-\Lambda_H}{\sigma_H}\right)\right)H(\bar{\tilde{k}}_\mu-\bar{\tilde{k}}_\mu^H) \nonumber \\
& & + \Phi_m(\infty)-1,
\label{cumultwosideddist}
\end{eqnarray}
where the quantities $\bar{\tilde{k}}_\mu^L, \bar{\tilde{k}}_\mu^H, \Lambda_y, \Lambda_L, \sigma_L, \Lambda_H, \sigma_H$ are defined as:
\begin{eqnarray}
 \bar{\tilde{k}}_\mu^L & = & \left(\frac{\mu-\mu_L}{\sigma}\right)^2, \label{eq:first}\\
 \Lambda_L & = & \frac{(\mu-\mu_L)(\mu+\mu_L-2\mu')}{\sigma^2}, \\
 \sigma_L & = & \frac{2(\mu-\mu_L)}{\sigma}, \\
 \bar{\tilde{k}}_\mu^H & = & \left(\frac{\mu_H-\mu}{\sigma}\right)^2, \\
 \Lambda_H & = & \frac{(\mu-\mu_H)(\mu+\mu_H-2\mu')}{\sigma^2}, \\
 \sigma_H & = & \frac{2(\mu_H-\mu)}{\sigma}, \\
 \Lambda_y & = & \frac{\mu-\mu'}{\sigma}, \label{eq:last}
\end{eqnarray}
the \emph{masked Gaussian integral} $\Phi_m(x)$ is defined as:
\begin{equation}
 \Phi_m(a) \equiv \int^{a}_{-\infty}\!\frac{1}{\sqrt{2\pi}}\exp\left[-\frac{x^2}{2}\right]m_\mu(x\sigma+\mu')\,\mathrm{d}\,x,
\label{eq:maskedgauss}
\end{equation}
and $H(x)$ is the discrete Heaviside function:
\begin{equation}
 H(x) = \begin{cases} 0 & x<0, \\ 1 & x\geq0. \end{cases}
\end{equation}
The pdf is given by:

\begin{equation}
\boxed{
\begin{array}{lcl}
  f(\bar{\tilde{k}}_\mu|\mu') & = & \Phi_m(\infty)\delta(\bar{\tilde{k}}_\mu) \\
& & + \frac{1}{2}\frac{1}{\sqrt{2\pi}}\frac{1}{\sqrt{\bar{\tilde{k}}_\mu}}\exp\left[-\frac{1}{2}\left(\sqrt{\bar{\tilde{k}}_\mu}-\Lambda_y\right)^2\right]   n_\mu\left(\Lambda_y - \sqrt{\bar{\tilde{k}}_\mu}\right)(1-H(\bar{\tilde{k}}_\mu-\bar{\tilde{k}}_\mu^L)) \\
& & + \frac{1}{2}\frac{1}{\sqrt{2\pi}}\frac{1}{\sqrt{\bar{\tilde{k}}_\mu}}\exp\left[-\frac{1}{2}\left(\sqrt{\bar{\tilde{k}}_\mu}+\Lambda_y\right)^2\right]   n_\mu\left(\Lambda_y + \sqrt{\bar{\tilde{k}}_\mu}\right)(1-H(\bar{\tilde{k}}_\mu-\bar{\tilde{k}}_\mu^H)) \\
& & + \frac{1}{\sigma_L\sqrt{2\pi}}\exp\left[-\frac{1}{2}\frac{\left(\bar{\tilde{k}}_\mu-\Lambda_L\right)^2}{\sigma_L^2}\right]n_\mu\left(\frac{\Lambda_L-\bar{\tilde{k}}_\mu}{\sigma_L}\right)H(\bar{\tilde{k}}_\mu-\bar{\tilde{k}}_\mu^L) \\ 
& & +
 \frac{1}{\sigma_H\sqrt{2\pi}}\exp\left[-\frac{1}{2}\frac{\left(\bar{\tilde{k}}_\mu-\Lambda_H\right)^2}{\sigma_H^2}\right]n_\mu\left(\frac{\bar{\tilde{k}}_\mu-\Lambda_H}{\sigma_H}\right)H(\bar{\tilde{k}}_\mu-\bar{\tilde{k}}_\mu^H),
\end{array}
}\label{twosideddist}
\end{equation}
where 
\begin{equation}
 n_\mu(x) = 1 - m_\mu(x\sigma + \mu').
\end{equation}

\subsection{Special Cases of the Test Statistic}
The likelihood-based test statistics that were originally described in~\cite{asymptoticpaper} and~\cite{asymptoticpaper2} can be considered as special cases of the general double-bounded custom-sided test statistic.  

\subsubsection{Distribution of $t_\mu$}
The unbounded two-sided test statistic $t_\mu$ is defined by:
\begin{eqnarray}
 \mu_L & = & -\infty, \\ 
 \mu_H & = & \infty, \\
 m_\mu(\hat\mu) & = & 0 \forall \hat\mu.
\end{eqnarray}
It has the following properties:
\begin{eqnarray}
 \tilde{q}_\mu^L & = & \infty, \\
 \tilde{q}_\mu^H & = & \infty \\
 \Phi_m(\infty) & = & 0 \\
 n_\mu(x) & = & 1 \forall x.
\end{eqnarray}
Therefore the pdf of $t_\mu$ is given by:
\begin{equation}
 \boxed{f(t_\mu|\mu') = \frac{1}{2}\frac{1}{\sqrt{2\pi}}\frac{1}{\sqrt{\tilde{t}_\mu}}\exp\left[-\frac{1}{2}\left(\sqrt{\tilde{t}_\mu}+\Lambda_y\right)^2\right] + \frac{1}{2}\frac{1}{\sqrt{2\pi}}\frac{1}{\sqrt{\tilde{t}_\mu}}\exp\left[-\frac{1}{2}\left(\sqrt{\tilde{t}_\mu}-\Lambda_y\right)^2\right].}
\end{equation}

\subsubsection{Distribution of $\tilde{t}_\mu$}
The lower bounded two-sided test statistic $\tilde{t}_\mu$ is defined by:
\begin{eqnarray}
 \mu_L & = & 0, \\ 
 \mu_H & = & \infty \\
 m_\mu(\hat\mu) & = & 0 \forall \hat\mu,
\end{eqnarray}
where $\mu_L$ has been chosen as 0 for consistency with~\cite{asymptoticpaper}. It has the following properties:
\begin{eqnarray}
 \tilde{q}_\mu^L & = & \mu^2/\sigma^2, \\
 \tilde{q}_\mu^H & = & \infty \\
 \Lambda_L & = & (\mu^2-2\mu\mu')/\sigma^2 \\
 \sigma_L & = & 2\mu/\sigma \\
 \Phi_m(\infty) & = & 0 \\
 n_\mu(x) & = & 1 \forall x.
\end{eqnarray}
Therefore the pdf of $\tilde{t}_\mu$ is given by:
\begin{equation}
\boxed{\begin{array}{lcl}f(\tilde{t}_\mu|\mu') = \begin{cases} 
                               \frac{1}{2}\frac{1}{\sqrt{2\pi}}\frac{1}{\sqrt{\tilde{t}_\mu}}\exp\left[-\frac{1}{2}\left(\sqrt{\tilde{t}_\mu}+\Lambda_y\right)^2\right] + \frac{1}{2}\frac{1}{\sqrt{2\pi}}\frac{1}{\sqrt{\tilde{t}_\mu}}\exp\left[-\frac{1}{2}\left(\sqrt{\tilde{t}_\mu}-\Lambda_y\right)^2\right] & \tilde{t}_\mu<\mu^2/\sigma^2, \\
                               \frac{1}{2}\frac{1}{\sqrt{2\pi}}\frac{1}{\sqrt{\tilde{t}_\mu}}\exp\left[-\frac{1}{2}\left(\sqrt{\tilde{t}_\mu}+\Lambda_y\right)^2\right] + \frac{1}{\sqrt{2\pi}(2\mu/\sigma)}\exp\left[-\frac{1}{2}\frac{\left(\tilde{t}_\mu-\frac{\mu^2-2\mu\mu'}{\sigma^2}\right)^2}{(s\mu/\sigma)^2}\right] & \mu^2/\sigma^2<\tilde{t}_\mu. \\
                        \end{cases}\end{array}}
\label{eq:f}
\end{equation}

\subsubsection{Distribution of $q_\mu$}
The unbounded one-sided test statistic $q_\mu$ is defined by:
\begin{eqnarray}
 \mu_L & = & -\infty, \\ 
 \mu_H & = & \infty \\
 m_\mu(\hat\mu) & = & \begin{cases} 
                   1 & \hat\mu>\mu,\\
                   0 & \hat\mu<\mu,
                  \end{cases}.
\end{eqnarray}
It has the following properties:
\begin{eqnarray}
 q_\mu^L & = & \infty, \\
 q_\mu^H & = & \infty \\
 \Phi_m(\infty) & = & \Phi\left(\frac{\mu-\mu'}{\sigma}\right).
\end{eqnarray}
One sees that:
\begin{eqnarray}
  (\Lambda_y + \sqrt{q_\mu})\sigma+\mu' & > & (\Lambda_y)\sigma+\mu' = \mu. \\
  (\Lambda_y - \sqrt{q_\mu})\sigma+\mu' & < & (\Lambda_y)\sigma+\mu' = \mu.
\end{eqnarray}
Therefore:
\begin{eqnarray}
 n_\mu\left(\Lambda_y + \sqrt{q_\mu}\right) & = & 0 \\
 n_\mu\left(\Lambda_y - \sqrt{q_\mu}\right) & = & 1,
\end{eqnarray}
leading to the pdf of $q_\mu$ being given by:
\begin{equation}
\boxed{
f(\tilde{q}_\mu|\mu') = 
                               \Phi\left(\frac{\mu-\mu'}{\sigma}\right)\delta(q_\mu) + 
\frac{1}{2}\frac{1}{\sqrt{2\pi}}\frac{1}{\sqrt{q_\mu}}\exp\left[-\frac{1}{2}\left(\sqrt{q_\mu}-\Lambda_y\right)^2\right]} 
\end{equation}

\subsubsection{Distribution of $\tilde{q}_\mu$}
The lower bounded one-sided test statistic $\tilde{q}_\mu$ is defined by:
\begin{eqnarray}
 \mu_L & = & 0, \\ 
 \mu_H & = & \infty \\
 m_\mu(\hat\mu) & = & \begin{cases} 
                   1 & \hat\mu>\mu,\\
                   0 & \hat\mu<\mu,
                  \end{cases}.
\end{eqnarray}
It has the following properties:
\begin{eqnarray}
 \tilde{q}_\mu^L & = & \mu^2/\sigma^2, \\
 \tilde{q}_\mu^H & = & \infty \\
 \Lambda_L & = & (\mu^2-2\mu\mu')/\sigma^2 \\
 \sigma_L & = & 2\mu/\sigma \\
 \Phi_m(\infty) & = & \Phi\left(\frac{\mu-\mu'}{\sigma}\right).
\end{eqnarray}
One sees that:
\begin{eqnarray}
  (\Lambda_y + \sqrt{\tilde{q}_\mu})\sigma+\mu' & > & (\Lambda_y)\sigma+\mu' = \mu. \\
  (\Lambda_y - \sqrt{\tilde{q}_\mu})\sigma+\mu' & < & (\Lambda_y)\sigma+\mu' = \mu.
\end{eqnarray}
Therefore:
\begin{eqnarray*}
 n_\mu\left(\Lambda_y + \sqrt{\tilde{q}_\mu}\right) & = & 0 \\
 n_\mu\left(\Lambda_y - \sqrt{\tilde{q}_\mu}\right) & = & 1
\end{eqnarray*}
Similarly, for $\tilde{q}_\mu^L<\tilde{q}_\mu$ one finds:
\begin{equation*}
\left(\frac{\Lambda_L - \tilde{q}_\mu}{\sigma_L}\right)\sigma+\mu'  < \left(\frac{\Lambda_L - \tilde{q}_\mu^L}{\sigma_L}\right)\sigma+\mu' = \mu_L = 0 < \mu 
\end{equation*}
Therefore:
\begin{eqnarray}
 n_\mu\left(\frac{\Lambda_L - \tilde{q}_\mu}{\sigma_L}\right) & = & 1,
\end{eqnarray}
leading to the pdf of $\tilde{q}_\mu$ being given by:
\begin{equation}
\boxed{
f(\tilde{q}_\mu|\mu') = \begin{cases} 
                               \Phi\left(\frac{\mu-\mu'}{\sigma}\right)\delta(\tilde{q}_\mu) + 
\frac{1}{2}\frac{1}{\sqrt{2\pi}}\frac{1}{\sqrt{\tilde{q}_\mu}}\exp\left[-\frac{1}{2}\left(\sqrt{\tilde{q}_\mu}-\Lambda_y\right)^2\right] & \tilde{q}_\mu<\mu^2/\sigma^2, \\
                               \Phi\left(\frac{\mu-\mu'}{\sigma}\right)\delta(\tilde{q}_\mu) + 
\frac{1}{\sigma_L\sqrt{2\pi}}\exp\left[-\frac{1}{2}\frac{\left(\tilde{q}_\mu-\Lambda_L\right)^2}{\sigma_L^2}\right]
 & \tilde{q}_\mu>\mu^2/\sigma^2. 
      \end{cases}} 
\end{equation}

\subsubsection{Distribution of $\bar{\tilde{t}}_\mu$}
The double-bounded two-sided test statistic $\bar{\tilde{t}}_\mu$ is defined by:
\begin{equation}
 m_\mu(\hat\mu) = 0.
\end{equation}
Thus all terms involving $m_\mu$ are zero, and the pdf becomes:
\begin{equation}
\boxed{\begin{array}{lcl}
 f(\bar{\tilde{t}}_\mu|\mu') & = & \frac{1}{2}\frac{1}{\sqrt{2\pi}}\frac{1}{\sqrt{\bar{\tilde{t}}_\mu}}\exp\left[-\frac{1}{2}\left(\sqrt{\bar{\tilde{t}}_\mu}-\Lambda_y\right)^2\right]   (1-H(\bar{\tilde{t}}_\mu-\bar{\tilde{t}}_\mu^L)) \\
& & + \frac{1}{2}\frac{1}{\sqrt{2\pi}}\frac{1}{\sqrt{\bar{\tilde{t}}_\mu}}\exp\left[-\frac{1}{2}\left(\sqrt{\bar{\tilde{t}}_\mu}+\Lambda_y\right)^2\right]   (1-H(\bar{\tilde{t}}_\mu-\bar{\tilde{t}}_\mu^H)) \\
& & + \frac{1}{\sigma_L\sqrt{2\pi}}\exp\left[-\frac{1}{2}\frac{\left(\bar{\tilde{t}}_\mu-\Lambda_L\right)^2}{\sigma_L^2}\right] H(\bar{\tilde{t}}_\mu-\bar{\tilde{t}}_\mu^L) \\ 
& & +
 \frac{1}{\sigma_H\sqrt{2\pi}}\exp\left[-\frac{1}{2}\frac{\left(\bar{\tilde{t}}_\mu-\Lambda_H\right)^2}{\sigma_H^2}\right] H(\bar{\tilde{t}}_\mu-\bar{\tilde{t}}_\mu^H),\end{array}}
\end{equation}
where the quantities $\bar{\tilde{t}}_\mu^L,\Lambda_L,\sigma_L,\bar{\tilde{t}}_\mu^H,\Lambda_H,\sigma_H$ and $\Lambda_y$ follow those given in equations~\ref{eq:first} to~\ref{eq:last}. This expression is compatible with the one presented in~\cite{asymptoticpaper2} up to a couple of sign differences present in the final term, which this author believes are nothing more than unintentional typographical errors present in the result of~\cite{asymptoticpaper2}.

\section{Toy MC Study}
\label{sec:example}
The accuracy of the asymptotic formula for the $\bar{\tilde{k}}_\mu$ distribution was tested with the following example.

One imagines a hypothetical experiment where a count of a number of events $n$ is made, where it is assumed that this count is modelled by a Poisson distribution with expectation given by $E[n]=\mu L\epsilon + b$. Here we take $b$ to be some background contribution, which will be constrained a Gaussian term. $L$ could represent a luminosity measurement, and $\epsilon$ may represent some acceptance for a hypothesised signal, which will have a cross-section given by $\mu$. $L$ and $\epsilon$ will also be constrained by Gaussian terms, such that if one were to observe $n$ events, the likelihood function for this observed data would be:
\begin{equation}
 L(\mu,b,L,\epsilon) = P(n|\mu L\epsilon + b)G(b_0|b,\sigma_b)G(L_0|L,\sigma_L)G(\epsilon_0|\epsilon,\sigma_\epsilon),
\end{equation}
where $b_0,L_0$ and $\epsilon_0$ represent auxiliary measurements that constrain the nuisance parameters, and $\sigma_b,\sigma_L$ and $\sigma_\epsilon$ are the corresponding uncertainties on those auxiliary measurements. $P(n|x)$ is the Poisson distribution with mean $x$, evaluated at $n$, and $G(x|a,b)$ is a Gaussian distribution with mean $a$ and standard deviation $b$, evaluated at $x$. 

Suppose, for the purposes of demonstration, that $\mu$ was believed to lie somewhere between -5 and 20, and for a null hypothesis was believed to be equal to 4. One seeks to assess deviations from this null hypothesis and set confidence limits on the parameter of interest $\mu$. Therefore one defines the test statistic $\bar{\tilde{k}}_\mu$ with the following parameters:

\begin{eqnarray}
 \mu_L & = & -5, \\ 
 \mu_H & = & 20 \\
 m_\mu(\hat\mu) & = & \begin{cases} 
                   1 & \mu<4,\hat\mu<\mu,\\
                   1 & \mu>4, \hat\mu>\mu,\\
                   0 & \text{otherwise}
                  \end{cases}.
\end{eqnarray}

The compatibility function is defined such that when testing a value of $\mu$ less than the null value of 4, data corresponding to values of $\hat\mu<\mu$ are considered fully compatible with the hypothesis of $\mu$. But for testing values of $\mu>4$, only data corresponding to values of $\hat\mu>\mu$ are considered compatible. 

For the example, values for the global observables ($b_0,L_0$ and $\epsilon_0$) and their uncertainties were somewhat arbitrarily chosen such that $\mu=-5$ corresponded to $(\mu L\epsilon + b)\approx24$ and $\mu=20$ corresponded to $(\mu L\epsilon + b)\approx113$, therefore in all cases of possible true values of the parameter of interest $\mu'$ the expected number of events is positive and not close to 0. For completeness, the nuisance parameter values used for generating pseudo-data for the global observables ($b_0,L_0$ and $\epsilon_0$) were taken as the conditional maximum likelihood estimator values, under the respective $\mu'$ hypothesis, using a data sample of 61 observed events. The standard deviation of parameter of interest estimator ($\sigma$) is estimated from the Asimov dataset, as discussed in~\cite{asymptoticpaper}.

\begin{figure}[!th]
 \begin{center}
	\subfigure[$\mu'=0$]{\includegraphics[width=0.47\textwidth]{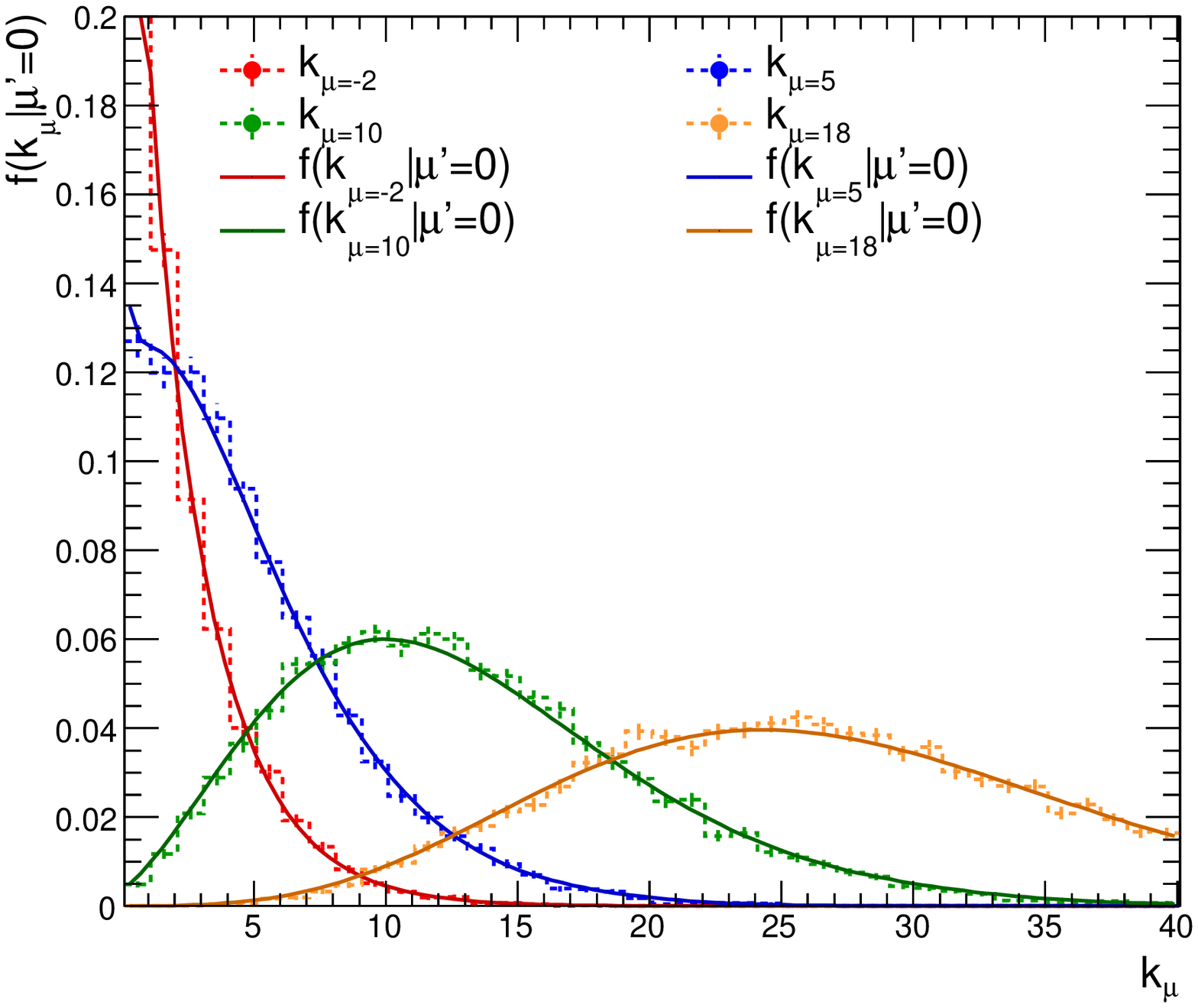}}
        \subfigure[$\mu'=4$]{\includegraphics[width=0.47\textwidth]{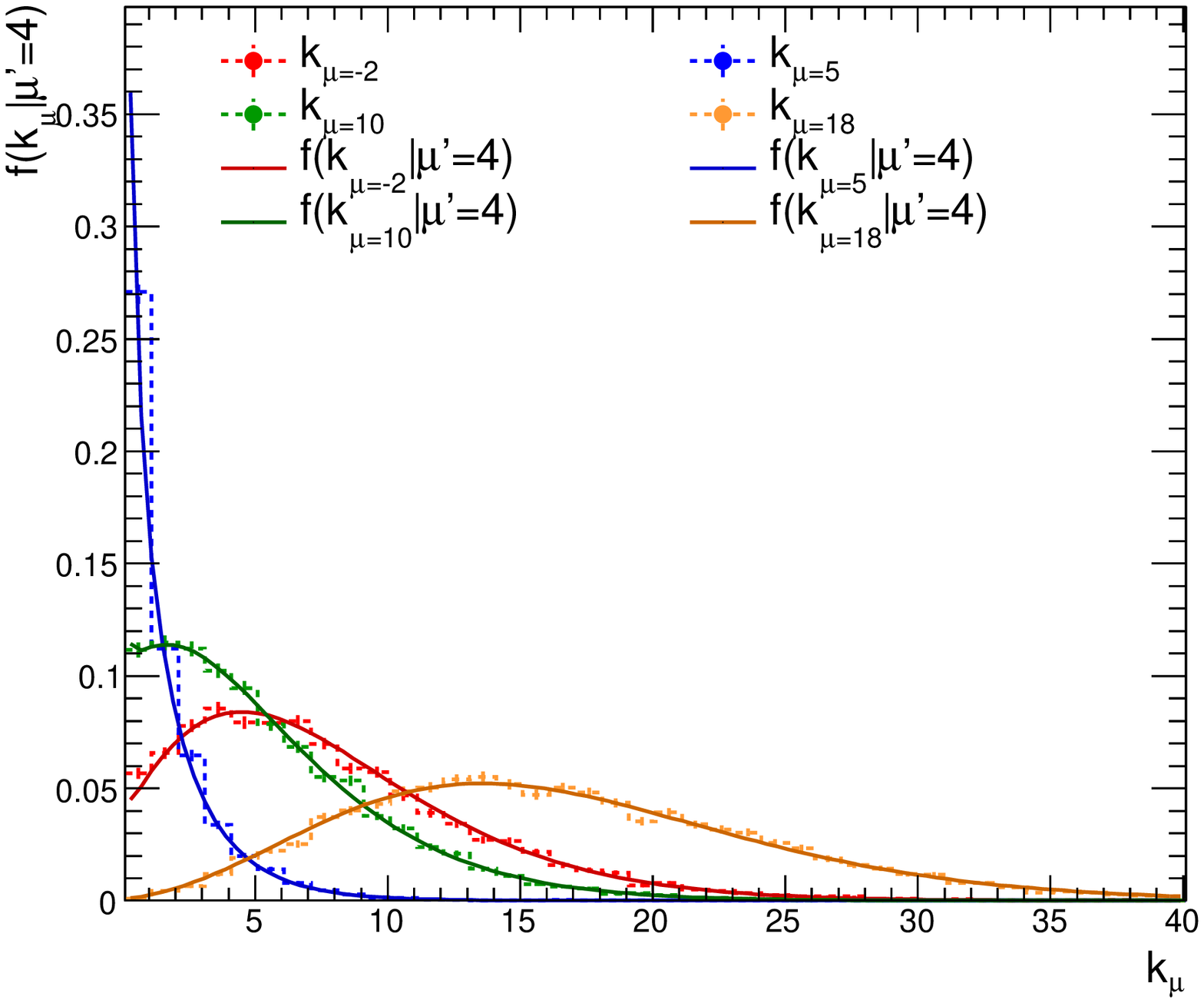}}
        \subfigure[$\mu'=8$]{\includegraphics[width=0.47\textwidth]{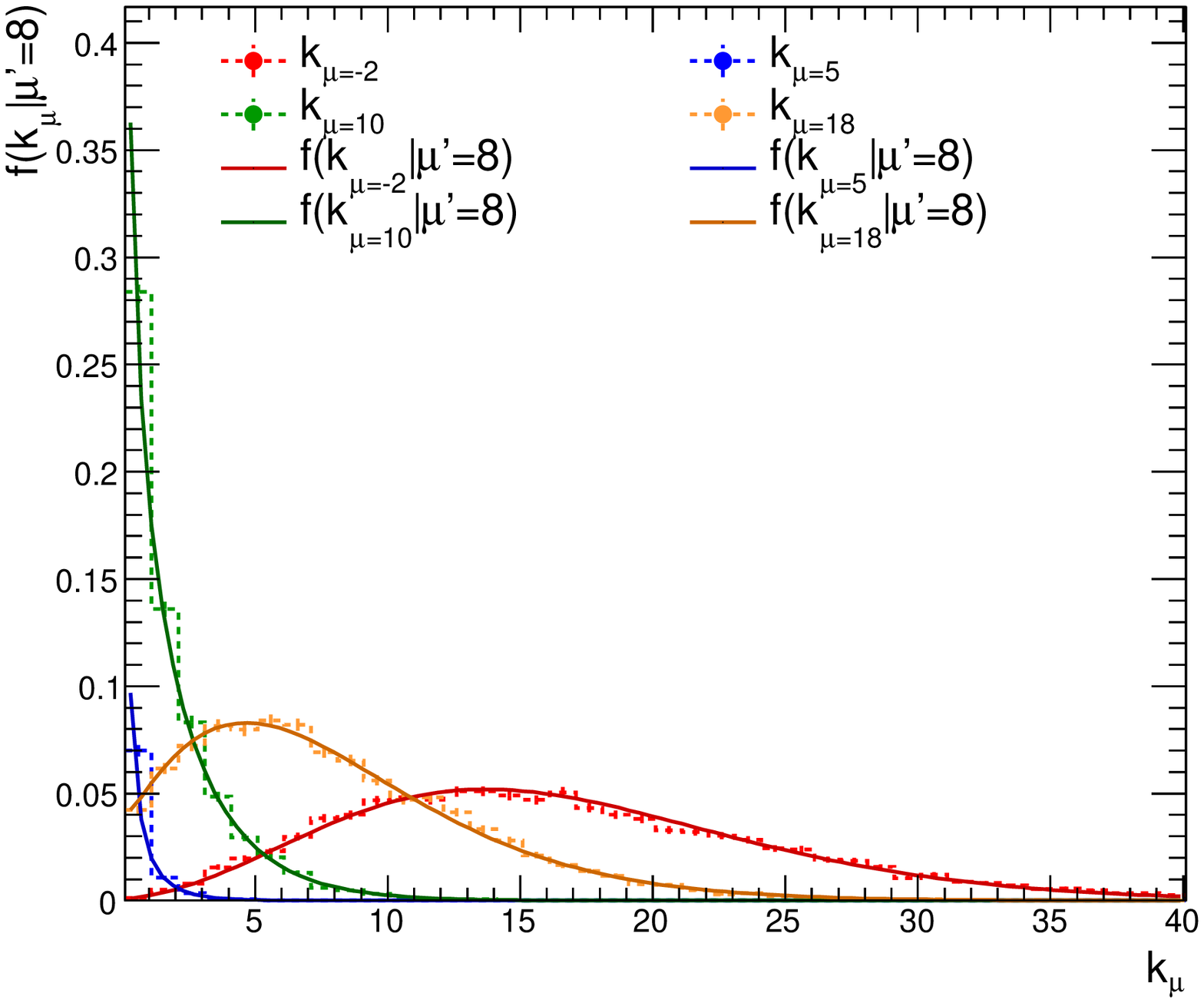}}
        \subfigure[$\mu'=18$]{\includegraphics[width=0.47\textwidth]{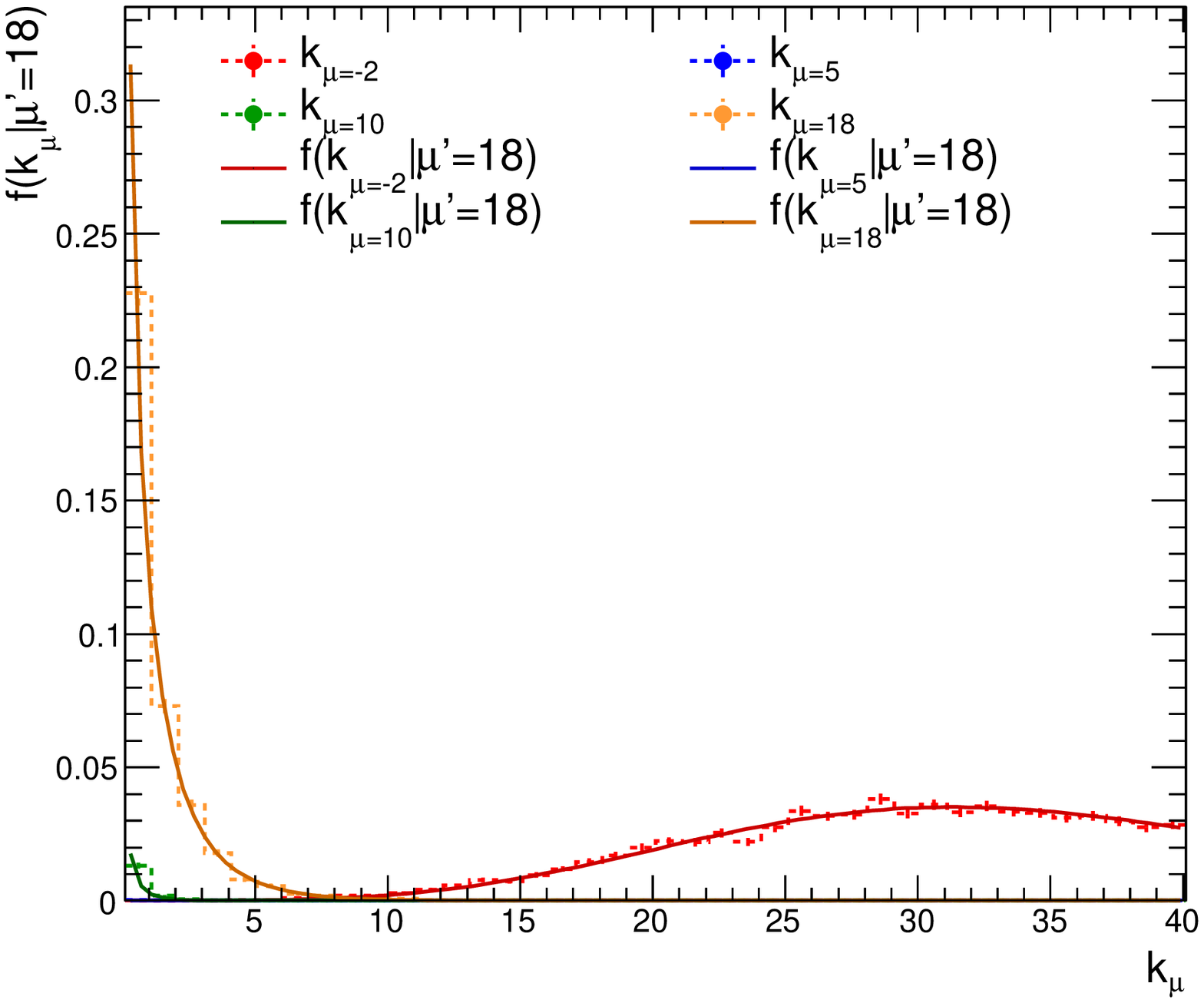}
        \label{fig:l1calo_fcal_stdev} }
        \caption{Distributions of the $\bar{\tilde{k}}_\mu$ test statistic with $\mu$ at values of -2,5,10 and 18, for different true values of the parameter of interest $\mu'$. }
        \label{fig:toymc}
    \end{center}
\end{figure}

Figure~\ref{fig:toymc} shows distributions of the test statistic $\bar{\tilde{k}}_\mu$ for a selection of values of $\mu$, under the hypothesis of four different true values of the parameter of interest $\mu'$. The distributions were generated using a Monte Carlo technique and are shown as dashed lines. The distributions obtained from the asymptotic formula given in equation~\ref{eq:f} are shown as solid curves. 

\vfill

\section{Derivation of $\bar{\tilde{k}}_\mu$ Distribution}
\label{sec:derivation}

One seeks the sampling distribution for the test statistic defined in equation~\ref{eq:teststatdef}. The method used is to determine the cumulative probability distribution under the Wald approximation, and then differentiate to arrive at the pdf.

The Wald approximation for a single parameter of interest states that:

\begin{equation}
 -2\ln{\frac{L(\mu,\hat{\hat{\theta}}(\mu))}{L(\hat{\mu},\hat{\theta})}} \approx \frac{(\mu-\hat{\mu})^2}{\sigma^2},
\label{waldapprox}
\end{equation}
which is valid in the case of large data sample sizes. Under the assumption of a particular true but unknown value of the signal strength parameter, $\mu'$, the maximum likelihood estimator of the signal strength, $\hat{\mu}$ follows a Gaussian distribution with mean $\mu'$ and standard deviation $\sigma$. For later convenience, this can be expressed as
\begin{equation}
 \hat{\mu} = \sigma x + \mu',
\label{xdef}
\end{equation}
where $x$ is a normally distributed random variable. 

Using the approximation~\ref{waldapprox}, one is able to write:

\begin{equation}
 \bar{\tilde{k}}_\mu = \begin{cases} 
  0 & m_\mu(\hat{\mu})=1, \\
  \frac{\mu^2-\mu_L^2-2(\mu-\mu_L)\hat{\mu}}{\sigma^2} & m_\mu(\hat{\mu})=0,\hat{\mu}<\mu_L,\\
  \frac{(\mu-\hat{\mu})^2}{\sigma^2} & m_\mu(\hat{\mu})=0,\mu_L<\hat{\mu}<\mu_H,\\
  \frac{\mu^2-\mu_H^2-2(\mu-\mu_H)\hat{\mu}}{\sigma^2} & m_\mu(\hat{\mu})=0,\hat{\mu}>\mu_H,
\end{cases}
\label{tildekapprox}
\end{equation}
where the expressions in the second and last cases come from expressing the associated likelihood ratio as the sum of two likelihood ratios, before apply the approximation~\ref{waldapprox}, for example:
\begin{equation}
\begin{array}{lcl|} -2\ln{\frac{L(\mu,\hat{\hat{\theta}}(\mu))}{L(\mu_L,\hat{\hat{\theta}}(\mu_L))}} & = & \left(-2\ln{\frac{L(\mu,\hat{\hat{\theta}}(\mu))}{L(\hat{\mu},\hat{\theta})}}\right) - \left(-2\ln{\frac{L(\mu_L,\hat{\hat{\theta}}(\mu_L))}{L(\hat{\mu},\hat{\theta})}}\right) \\ & \approx & \frac{(\mu-\hat{\mu})^2}{\sigma^2} - \frac{(\mu_L-\hat{\mu})^2}{\sigma^2}.
\end{array}
\end{equation}

The possible values of $\bar{\tilde{k}}_\mu$ range from 0 to $\infty$, but one finds that each region of $\hat{\mu}$ produces values of $\bar{\tilde{k}}_\mu$ in certain ranges. For the case $\hat{\mu}< \mu_L$, because $(\mu-\mu_L)\geq 0$ (one assumes that you would not wish to construct this test statistic for a value of $\mu$ outside of the physically allowed region), we find that this range of $\hat{\mu}$ produce $\bar{\tilde{k}}_\mu$ greater than a minimum value. A similar argument applies to $\hat{\mu}> \mu_H$, where $(\mu-\mu_H)\leq 0$:
\begin{equation}
 \bar{\tilde{k}}_\mu \geq \begin{cases}
                           \frac{\mu^2-\mu_L^2-2(\mu-\mu_L)\mu_L}{\sigma^2} = \left(\frac{\mu-\mu_L}{\sigma}\right)^2 \equiv \bar{\tilde{k}}_\mu^L & m_\mu(\hat{\mu})=0,\hat{\mu}<\mu_L,\\
                           \frac{\mu^2-\mu_L^2-2(\mu-\mu_H)\mu_H}{\sigma^2} = \left(\frac{\mu-\mu_H}{\sigma}\right)^2 \equiv \bar{\tilde{k}}_\mu^H & m_\mu(\hat{\mu})=0,\hat{\mu}>\mu_H,
                          \end{cases} 
\end{equation}
where $\bar{\tilde{k}}_\mu^L$ and $\bar{\tilde{k}}_\mu^H$ are introduced as convenient labels for the values defining the boundaries of each of these regions. The order of these limits defines two cases that must be treated separately. We start with the case where $|\mu-\mu_L|<|\mu-\mu_H|$, i.e. where one has formed the test statistic for a value of $\mu$ that is closer to the lower bound than it is to the upper bound.

\subsection{Case where $\bar{\tilde{k}}_\mu^L<\bar{\tilde{k}}_\mu^H$}
\label{sec:case1}
For the cases where $m_\mu(\hat{\mu})=0$, three regions of possible $\bar{\tilde{k}}_\mu$ are defined with the associated possible value of $\hat\mu$:
\begin{equation}
 \begin{array}{lcl|} 0 \leq \bar{\tilde{k}}_\mu < \bar{\tilde{k}}_\mu^L & : & \mu_L<\hat{\mu}<\mu_H, \\
  \bar{\tilde{k}}_\mu^L \leq \bar{\tilde{k}}_\mu < \bar{\tilde{k}}_\mu^H & : & \hat{\mu}<\mu_H, \\
  \bar{\tilde{k}}_\mu^H \leq \bar{\tilde{k}}_\mu & : & \hat{\mu}<\mu_L, \hat{\mu}>\mu_H,
 \end{array}
\end{equation}
One can construct the cumulative probability distribution for each region of $\bar{\tilde{k}}_\mu$ values. We start with lowest value region.

\subsubsection{$0 \leq \bar{\tilde{k}}_\mu < \bar{\tilde{k}}_\mu^L$}
The contributions to this region are the cases where $m_\mu(\hat{\mu})=1$ and $(m_\mu(\hat{\mu})=0,\mu_L<\hat{\mu}<\mu_H)$. One defines the quantity $y$:
\begin{equation}
 y = \frac{\mu-\hat{\mu}}{\sigma},
\label{ydef}
\end{equation}
such that $\bar{\tilde{k}}_\mu=y^2$ (this is true in this region because the other two expressions for $\bar{\tilde{k}}_\mu$ as given in equation \ref{tildekapprox} correspond to values of $\hat{\mu}$ that do not contribute to this range of $\bar{\tilde{k}}_\mu$ values). Therefore to obtain a value of the test statistic less than $\bar{\tilde{k}}_\mu$ corresponds to either $m_\mu(\hat{\mu})=1$, or $m_\mu(\hat{\mu})=0$ and a value of $y$ in the range:
\begin{equation}
 -\sqrt{\bar{\tilde{k}}_\mu} < y < \sqrt{\bar{\tilde{k}}_\mu}.
\end{equation}
Using the distribution of $\hat{\mu}$ (equation~\ref{xdef}), we can deduce that $y$ is also a Gaussian distributed quantity, with mean and standard deviation given by:
\begin{equation}
 \Lambda_y = \frac{\mu-\mu'}{\sigma},
 \sigma_y = 1.
\end{equation}
The cumulative distribution for $\bar{\tilde{k}}_\mu$ in this region is given by:
\begin{equation}
 F_1(\bar{\tilde{k}}_\mu|\mu') = P(m_\mu(\hat{\mu})=1) + P(-\sqrt{\bar{\tilde{k}}_\mu} < y < \sqrt{\bar{\tilde{k}}_\mu}, \mu_L<\hat{\mu}<\mu_H,m_\mu(\hat{\mu})=0).
\end{equation}
The constraint terms can be re-expressed in terms of the random variable $y$:
\begin{eqnarray}
 \mu_L<\hat{\mu}<\mu_H & \rightarrow & \frac{\mu-\mu_H}{\sigma}<y<\frac{\mu-\mu_L}{\sigma}, \label{eq:constraint1}\\
 m_\mu(\hat{\mu})=0 & \rightarrow & m_\mu(\mu-\sigma y)=0.
\end{eqnarray}
The first of these constraints has no effect in this region of $\bar{\tilde{k}}_\mu$ (we are only considering $\bar{\tilde{k}}_\mu<\bar{\tilde{k}}_\mu^L<\bar{\tilde{k}}_\mu^H$, therefore this constraint is always satisfied here). So the cumulative distribution can be expressed as:
\begin{equation}
 F_1(\bar{\tilde{k}}_\mu|\mu') = P(m_\mu(\hat{\mu})=1) + P(-\sqrt{\bar{\tilde{k}}_\mu} < y < \sqrt{\bar{\tilde{k}}_\mu}, m_\mu(\mu-\sigma y)=0).
\end{equation}
The second term corresponds to the probability for $y$ to lie between $-\sqrt{\bar{\tilde{k}}_\mu}$ and $\sqrt{\bar{\tilde{k}}_\mu}$, but where it must not result in $m_\mu(\mu-\sigma y)=1$. This is equivalent to the probability of $-\sqrt{\bar{\tilde{k}}_\mu}<y<\sqrt{\bar{\tilde{k}}_\mu}$, less the probability of both $-\sqrt{\bar{\tilde{k}}_\mu}<y<\sqrt{\bar{\tilde{k}}_\mu}$ and $m_\mu(\mu-\sigma y)=1$ being satisfied. Hence
\begin{equation}
 F_1(\bar{\tilde{k}}_\mu|\mu') = P(m_\mu(\hat{\mu})=1) + P(-\sqrt{\bar{\tilde{k}}_\mu} < y < \sqrt{\bar{\tilde{k}}_\mu}) - P(-\sqrt{\bar{\tilde{k}}_\mu} < y < \sqrt{\bar{\tilde{k}}_\mu}, m_\mu(\mu-\sigma y)=1).
\end{equation}
Due to the random variables $\hat{\mu}$ and $y$ being Gaussian distributed, the probabilities can be expressed as integrals:
\begin{eqnarray}
  F_1(\bar{\tilde{k}}_\mu|\mu') & = & \int^{\infty}_{-\infty}\!\frac{1}{\sigma\sqrt{2\pi}}\exp\left[\frac{(\hat{\mu}-\mu')^2}{2\sigma^2}\right]m_\mu(\hat\mu)\,\mathrm{d}\hat\mu \nonumber \\
& & + \Phi(\sqrt{\bar{\tilde{k}}_\mu}-\Lambda_y) - \Phi(-\sqrt{\bar{\tilde{k}}_\mu}-\Lambda_y) \nonumber \\
& & - \int^{\sqrt{\bar{\tilde{k}}_\mu}}_{-\sqrt{\bar{\tilde{k}}_\mu}}\!\frac{1}{\sqrt{2\pi}}\exp\left[\frac{(y-\Lambda_y)^2}{2}\right]m_\mu(\mu-\sigma y)\,\mathrm{d}\,y.
\end{eqnarray}
Equation~\ref{ydef} and ~\ref{xdef} are used to perform a change of variables, giving:
\begin{eqnarray}
F_1(\bar{\tilde{k}}_\mu|\mu') & = & \int^{\infty}_{-\infty}\!\frac{1}{\sqrt{2\pi}}\exp\left[-\frac{x^2}{2}\right]m_\mu(x\sigma+\mu')\,\mathrm{d}\,x \nonumber \\
& & + \Phi(\sqrt{\bar{\tilde{k}}_\mu}-\Lambda_y) - 1 + \Phi(\sqrt{\bar{\tilde{k}}_\mu}+\Lambda_y) \nonumber \\
& & - \int^{\Lambda_y+\sqrt{\bar{\tilde{k}}_\mu}}_{\Lambda_y-\sqrt{\bar{\tilde{k}}_\mu}}\!\frac{1}{\sqrt{2\pi}}\exp\left[-\frac{x^2}{2}\right]m_\mu(x\sigma+\mu')\,\mathrm{d}\,x,
\end{eqnarray}
where the follow useful relationship was also used:
\begin{equation}
 \Phi(x) = 1 - \Phi(-x).
\label{eq:useful}
\end{equation}
Using the \emph{masked Gaussian integral} defined in equation~\ref{eq:maskedgauss} the cumulative distribution is finally expressed as:
\begin{eqnarray}
F_1(\bar{\tilde{k}}_\mu|\mu') & = & \Phi_m(\infty) + \Phi_m(\Lambda_y-\sqrt{\bar{\tilde{k}}_\mu}) - \Phi_m(\Lambda_y+\sqrt{\bar{\tilde{k}}_\mu}) \nonumber \\
& & + \Phi(\sqrt{\bar{\tilde{k}}_\mu}-\Lambda_y) - 1 + \Phi(\sqrt{\bar{\tilde{k}}_\mu}+\Lambda_y).
\label{eq:F1_custom}
\end{eqnarray}

\subsubsection{$\bar{\tilde{k}}_\mu^L \leq \bar{\tilde{k}}_\mu < \bar{\tilde{k}}_\mu^H$}
The next region of $\bar{\tilde{k}}_\mu$ values has contributions from the $\hat{\mu}<\mu_L$ and $\mu_L< \hat{\mu}< \mu_H$ regions of possible $\hat{\mu}$ values (for the cases where $m_\mu(\hat\mu)=0$).

Define the random variable $z_L$, which is the value of $\bar{\tilde{k}}_\mu$ when $\hat\mu<\mu_L$:
\begin{eqnarray}
 z_L & = & \frac{\mu^2-\mu_L^2-2(\mu-\mu_L)\hat{\mu}}{\sigma^2} \nonumber \\
 & = & \frac{\mu^2-\mu_L^2-2(\mu-\mu_L)\mu'}{\sigma^2} - \frac{2(\mu-\mu_L)}{\sigma}x \nonumber \\
 & = & \Lambda_L - \sigma_L x, 
\label{zldef}
\end{eqnarray}
i.e. $z_L$ is Gaussian distribution with mean and standard deviation given by:
\begin{equation}
 \Lambda_L = \frac{\mu^2-\mu_L^2-2(\mu-\mu_L)\mu'}{\sigma^2},
 \sigma_L = \frac{2(\mu-\mu_L)}{\sigma}.
\end{equation}
For the $\mu_L< \hat{\mu}< \mu_H$ case, the constraint in equation~\ref{eq:constraint1} now comes into effect at the upper limit of $(\mu-\mu_L)/\sigma = \sqrt{\bar{\tilde{k}}_\mu^L}$. The cumulative distribution in this region is given by:
\begin{eqnarray}
F_2(\bar{\tilde{k}}_\mu|\mu') & = & \Phi_m(\infty) + \Phi_m(\Lambda_y - \sqrt{\bar{\tilde{k}}_\mu^L}) - \Phi_m(\Lambda_y+\sqrt{\bar{\tilde{k}}_\mu}) \nonumber \\
& & + \Phi(\sqrt{\bar{\tilde{k}}_\mu^L}-\Lambda_y) - 1 + \Phi(\sqrt{\bar{\tilde{k}}_\mu}+\Lambda_y) \nonumber \\
& & + P(m_\mu(\hat{\mu})=0)P(\hat{\mu}<\mu_L|m_\mu(\hat{\mu})=0)P(z_L < \bar{\tilde{k}}_\mu | m_\mu(\hat{\mu})=0,\hat{\mu}<\mu_L).
\end{eqnarray}
Re-expressing the constraints in the last line in terms of $z_L$, one has:
\begin{eqnarray}
 \hat{\mu}<\mu_L & \rightarrow & z_L > \bar{\tilde{k}}_\mu^L,\\
 m_\mu(\hat{\mu})=0 & \rightarrow & m_\mu\left(\frac{z_L\sigma^2 + \mu_L^2 - \mu^2}{2(\mu_L-\mu)}\right)=0.
\end{eqnarray}
The distribution therefore can be expressed as:
\begin{eqnarray}
F_2(\bar{\tilde{k}}_\mu|\mu') & = & \Phi_m(\infty) + \Phi_m(\Lambda_y - \sqrt{\bar{\tilde{k}}_\mu^L}) - \Phi_m(\Lambda_y+\sqrt{\bar{\tilde{k}}_\mu}) \nonumber \\
& & + \Phi(\sqrt{\bar{\tilde{k}}_\mu^L}-\Lambda_y) - 1 + \Phi(\sqrt{\bar{\tilde{k}}_\mu}+\Lambda_y) \nonumber \\
& & + P(\bar{\tilde{k}}_\mu^L < z_L < \bar{\tilde{k}}_\mu, m_\mu(\hat{\mu})=0),
\end{eqnarray}
which can be written in integral form as:
\begin{eqnarray}
F_2(\bar{\tilde{k}}_\mu|\mu') & = & \Phi_m(\infty) + \Phi_m(\Lambda_y - \sqrt{\bar{\tilde{k}}_\mu^L}) - \Phi_m(\Lambda_y+\sqrt{\bar{\tilde{k}}_\mu}) \nonumber \\
& & + \Phi(\sqrt{\bar{\tilde{k}}_\mu^L}-\Lambda_y) - 1 + \Phi(\sqrt{\bar{\tilde{k}}_\mu}+\Lambda_y) \nonumber \\
& & + \Phi\left(\frac{\bar{\tilde{k}}_\mu-\Lambda_L}{\sigma_L}\right) - \Phi\left(\frac{\bar{\tilde{k}}_\mu^L-\Lambda_L}{\sigma_L}\right) \nonumber \\
& & - \int^{\bar{\tilde{k}}_\mu}_{\bar{\tilde{k}}_\mu^L}\!\frac{1}{\sigma_L\sqrt{2\pi}}\exp\left[\frac{(z_L-\Lambda_L)^2}{2\sigma_L^2}\right]m_\mu\left(\frac{z_L\sigma^2 + \mu_L^2 - \mu^2}{2(\mu_L-\mu)}\right)\,\mathrm{d}\,z_L.
\end{eqnarray}
A change of variables using the definition of $z_L$ given in equation~\ref{zldef} gives:
\begin{eqnarray}
F_2(\bar{\tilde{k}}_\mu|\mu') & = & \Phi_m(\infty) + \Phi_m((\mu_L-\mu')/\sigma) - \Phi_m(\Lambda_y+\sqrt{\bar{\tilde{k}}_\mu}) \nonumber\\
& & + \Phi(\sqrt{\bar{\tilde{k}}_\mu^L}-\Lambda_y) - 1 + \Phi(\sqrt{\bar{\tilde{k}}_\mu}+\Lambda_y) \\
& & + \Phi\left(\frac{\bar{\tilde{k}}_\mu-\Lambda_L}{\sigma_L}\right) - \Phi\left(\frac{\bar{\tilde{k}}_\mu^L-\Lambda_L}{\sigma_L}\right) \nonumber\\
& & - \int^{(\Lambda_L-\bar{\tilde{k}}_\mu^L)/\sigma_L}_{(\Lambda_L-\bar{\tilde{k}}_\mu)/\sigma_L}\!\frac{1}{\sqrt{2\pi}}\exp\left[-\frac{x^2}{2}\right]m_\mu\left(x\sigma+\mu'\right)\,\mathrm{d}\,x. 
\end{eqnarray}
By noting that
\begin{eqnarray}
 \frac{\bar{\tilde{k}}_\mu^L-\Lambda_L}{\sigma_L} & = & \frac{(\mu-\mu_L)^2 - [(\mu+\mu_L)(\mu-\mu_L) - 2(\mu-\mu_L)\mu']}{2(\mu-\mu_L)\sigma} \nonumber\\
 & = & \frac{(\mu'-\mu_L)}{\sigma} \nonumber\\
 & = & \sqrt{\bar{\tilde{k}}_\mu^L} - \Lambda_y,
\label{eq:simpL}
\end{eqnarray}
the cumulative distribution is simplified to
\begin{eqnarray}
F_2(\bar{\tilde{k}}_\mu|\mu') & = & \Phi_m(\infty) - \Phi_m(\Lambda_y+\sqrt{\bar{\tilde{k}}_\mu}) - 1 + \Phi(\sqrt{\bar{\tilde{k}}_\mu}+\Lambda_y) \nonumber\\
& & + \Phi\left(\frac{\bar{\tilde{k}}_\mu-\Lambda_L}{\sigma_L}\right) + \Phi_m\left(\frac{\Lambda_L-\bar{\tilde{k}}_\mu}{\sigma_L}\right).
\end{eqnarray}

\subsubsection{$\bar{\tilde{k}}_\mu \geq \bar{\tilde{k}}_\mu^H$}
The two regions of possible $\hat{\mu}$ values contributing to this region of $\bar{\tilde{k}}_\mu$ are the $\hat{\mu}<\mu_L$ and $\hat{\mu}>\mu_H$. Introduce the random variable $z_H$ defined as:
\begin{eqnarray}
 z_H & = & \frac{\mu^2-\mu_H^2-2(\mu-\mu_H)\hat{\mu}}{\sigma^2} \nonumber\\
     & = & \frac{\mu^2-\mu_H^2-2(\mu-\mu_H)\mu'}{\sigma^2} + \frac{2(\mu_H-\mu)}{\sigma}x \nonumber\\
     & = & \Lambda_H + \sigma_H x,
\end{eqnarray}
i.e. $z_H$ is Gaussian distributed mean and standard deviation given by:
\begin{equation}
 \Lambda_H = \frac{\mu^2-\mu_H^2-2(\mu-\mu_H)\mu'}{\sigma^2},
 \sigma_H =  \frac{2(\mu_H-\mu)}{\sigma}.
\end{equation}
Both limits in equation~\ref{eq:constraint1} apply to contributions coming from $(m_\mu(\hat{\mu})=0,\mu_L<\hat{\mu}<\mu_H)$ (i.e. any time this case occurs will result in a test statistic value smaller than $\bar{\tilde{k}}_\mu$), and the constraints on $z_H$ are similar to the constraints on $z_L$ given in the last section:
\begin{eqnarray}
F_3(\bar{\tilde{k}}_\mu|\mu') & = & \Phi_m(\infty) - \Phi_m(\Lambda_y+\sqrt{\bar{\tilde{k}}_\mu^H}) - 1 + \Phi(\sqrt{\bar{\tilde{k}}_\mu^H}+\Lambda_y) + \Phi\left(\frac{\bar{\tilde{k}}_\mu-\Lambda_L}{\sigma_L}\right) + \Phi_m\left(\frac{\Lambda_L-\bar{\tilde{k}}_\mu}{\sigma_L}\right) \nonumber\\
& & + P(\bar{\tilde{k}}_\mu^H < z_H < \bar{\tilde{k}}_\mu, m_\mu(\hat{\mu})=0).
\end{eqnarray}
Note that:
\begin{eqnarray}
 \frac{\bar{\tilde{k}}_\mu^H-\Lambda_H}{\sigma_H} & = & \frac{(\mu-\mu_H)^2 - [(\mu+\mu_H)(\mu-\mu_H) - 2(\mu-\mu_H)\mu']}{2(\mu_H-\mu)\sigma} \nonumber\\
 & = & \frac{(\mu_H-\mu')}{\sigma} \nonumber\\
 & = & \sqrt{\bar{\tilde{k}}_\mu^H} + \Lambda_y.
\label{eq:simpH}
\end{eqnarray}
Using this and changing variables as before, one finds:
\begin{eqnarray}
F_3(\bar{\tilde{k}}_\mu|\mu') & = & \Phi_m(\infty) - \Phi_m(\Lambda_y+\sqrt{\bar{\tilde{k}}_\mu^H}) - 1 + \Phi(\sqrt{\bar{\tilde{k}}_\mu^H}+\Lambda_y) + \Phi\left(\frac{\bar{\tilde{k}}_\mu-\Lambda_L}{\sigma_L}\right) + \Phi_m\left(\frac{\Lambda_L-\bar{\tilde{k}}_\mu}{\sigma_L}\right) \nonumber\\
& & + \Phi\left(\frac{\bar{\tilde{k}}_\mu-\Lambda_H}{\sigma_H}\right) - \Phi\left(\frac{\bar{\tilde{k}}_\mu^H-\Lambda_H}{\sigma_H}\right) - \Phi_m\left(\frac{\bar{\tilde{k}}_\mu-\Lambda_H}{\sigma_H}\right) + \Phi_m\left(\frac{\bar{\tilde{k}}_\mu^H-\Lambda_H}{\sigma_H}\right) \nonumber\\
& = & \Phi_m(\infty) + \Phi\left(\frac{\bar{\tilde{k}}_\mu-\Lambda_L}{\sigma_L}\right) + \Phi_m\left(\frac{\Lambda_L-\bar{\tilde{k}}_\mu}{\sigma_L}\right) \nonumber\\
& & + \Phi\left(\frac{\bar{\tilde{k}}_\mu-\Lambda_H}{\sigma_H}\right) - \Phi_m\left(\frac{\bar{\tilde{k}}_\mu-\Lambda_H}{\sigma_H}\right) - 1.
\label{eq:F3_custom}
\end{eqnarray}

\subsection{Case where $\bar{\tilde{k}}_\mu^L \geq \bar{\tilde{k}}_\mu^H$}
In this case, the test statistic is being formed for a value of $\mu$ closer to the upper physical bound ($\mu_H$) than to the lower physical bound ($\mu_L$). The possible values of $\bar{\tilde{k}}_\mu$ are again divided into three regions, and the cumulative distribution determined for each. 

\subsubsection{$\bar{\tilde{k}}_\mu < \bar{\tilde{k}}_\mu^H$}
In this region, only the $\mu_L<\hat{\mu}<\mu_H$ values contribute, so again the cumulative distribution is found to be $F_1(\bar{\tilde{k}}_\mu|\mu')$, given in equation~\ref{eq:F1_custom}.

\subsubsection{$\bar{\tilde{k}}_\mu^H \leq \bar{\tilde{k}}_\mu < \bar{\tilde{k}}_\mu^L$}
The lower limit of the constraint in equation~\ref{eq:constraint1} is reached. The cumulative distribution is given by:
\begin{eqnarray}
F_2(\bar{\tilde{k}}_\mu|\mu') & = & \Phi_m(\infty) + \Phi_m(\Lambda_y - \sqrt{\bar{\tilde{k}}_\mu}) - \Phi_m(\Lambda_y+\sqrt{\bar{\tilde{k}}_\mu^H}) \nonumber\\
& & + \Phi(\sqrt{\bar{\tilde{k}}_\mu}-\Lambda_y) - 1 + \Phi(\sqrt{\bar{\tilde{k}}_\mu^H}+\Lambda_y) \nonumber\\
& & + P(\bar{\tilde{k}}_\mu^H < z_H < \bar{\tilde{k}}_\mu, m_\mu(\hat{\mu})=0) \nonumber\\
& = & \Phi_m(\infty) + \Phi_m(\Lambda_y - \sqrt{\bar{\tilde{k}}_\mu}) - \Phi_m(\Lambda_y+\sqrt{\bar{\tilde{k}}_\mu^H}) \nonumber\\
& & + \Phi(\sqrt{\bar{\tilde{k}}_\mu}-\Lambda_y) - 1 + \Phi(\sqrt{\bar{\tilde{k}}_\mu^H}+\Lambda_y) \nonumber\\
& & +  \Phi\left(\frac{\bar{\tilde{k}}_\mu-\Lambda_H}{\sigma_H}\right) - \Phi\left(\frac{\bar{\tilde{k}}_\mu^H-\Lambda_H}{\sigma_H}\right) \nonumber\\
& & - \Phi_m\left(\frac{\bar{\tilde{k}}_\mu-\Lambda_H}{\sigma_H}\right) + \Phi_m\left(\frac{\bar{\tilde{k}}_\mu^H-\Lambda_H}{\sigma_H}\right) \nonumber\\
& = & \Phi_m(\infty) + \Phi_m(\Lambda_y - \sqrt{\bar{\tilde{k}}_\mu}) + \Phi(\sqrt{\bar{\tilde{k}}_\mu}-\Lambda_y) - 1 \nonumber\\
& & +  \Phi\left(\frac{\bar{\tilde{k}}_\mu-\Lambda_H}{\sigma_H}\right) - \Phi_m\left(\frac{\bar{\tilde{k}}_\mu-\Lambda_H}{\sigma_H}\right)
\end{eqnarray}

\subsubsection{$\bar{\tilde{k}}_\mu \geq \bar{\tilde{k}}_\mu^L$}
Both limits of equation~\ref{eq:constraint1} are reached, and the cumulative distribution becomes identical to $F_3(\bar{\tilde{k}}_\mu|\mu')$ given in equation~\ref{eq:F3_custom}.

\subsection{Summary of distribution}
The cumulative distribution in full is given by:
\begin{equation}
F(\bar{\tilde{k}}_\mu|\mu') = \begin{cases} 
                               \Phi_m(\infty) + \Phi(\sqrt{\bar{\tilde{k}}_\mu}-\Lambda_y) + \Phi(\sqrt{\bar{\tilde{k}}_\mu}+\Lambda_y) - 1 + \Phi_m(\Lambda_y-\sqrt{\bar{\tilde{k}}_\mu}) - \Phi_m(\Lambda_y+\sqrt{\bar{\tilde{k}}_\mu}) & \bar{\tilde{k}}_\mu<\bar{\tilde{k}}_\mu^L<\bar{\tilde{k}}_\mu^H, \\
                               \Phi_m(\infty) + \Phi(\sqrt{\bar{\tilde{k}}_\mu}+\Lambda_y) + \Phi\left(\frac{\bar{\tilde{k}}_\mu-\Lambda_L}{\sigma_L}\right) - 1 + \Phi_m\left(\frac{\Lambda_L-\bar{\tilde{k}}_\mu}{\sigma_L}\right) - \Phi_m(\Lambda_y+\sqrt{\bar{\tilde{k}}_\mu}) & \bar{\tilde{k}}_\mu^L<\bar{\tilde{k}}_\mu<\bar{\tilde{k}}_\mu^H, \\
                               \Phi_m(\infty) + \Phi\left(\frac{\bar{\tilde{k}}_\mu-\Lambda_L}{\sigma_L}\right) + \Phi\left(\frac{\bar{\tilde{k}}_\mu-\Lambda_H}{\sigma_H}\right) - 1 + \Phi_m\left(\frac{\Lambda_L-\bar{\tilde{k}}_\mu}{\sigma_L}\right) - \Phi_m\left(\frac{\bar{\tilde{k}}_\mu-\Lambda_H}{\sigma_H}\right) & \bar{\tilde{k}}_\mu^L<\bar{\tilde{k}}_\mu^H<\bar{\tilde{k}}_\mu \\ 
                               \Phi_m(\infty) + \Phi(\sqrt{\bar{\tilde{k}}_\mu}-\Lambda_y) + \Phi(\sqrt{\bar{\tilde{k}}_\mu}+\Lambda_y) - 1 + \Phi_m(\Lambda_y-\sqrt{\bar{\tilde{k}}_\mu}) - \Phi_m(\Lambda_y+\sqrt{\bar{\tilde{k}}_\mu}) & \bar{\tilde{k}}_\mu<\bar{\tilde{k}}_\mu^H<\bar{\tilde{k}}_\mu^L, \\
                               \Phi_m(\infty) + \Phi(\sqrt{\bar{\tilde{k}}_\mu}-\Lambda_y) +  \Phi\left(\frac{\bar{\tilde{k}}_\mu-\Lambda_H}{\sigma_H}\right) - 1 + \Phi_m(\Lambda_y - \sqrt{\bar{\tilde{k}}_\mu}) - \Phi_m\left(\frac{\bar{\tilde{k}}_\mu-\Lambda_H}{\sigma_H}\right) & \bar{\tilde{k}}_\mu^H<\bar{\tilde{k}}_\mu<\bar{\tilde{k}}_\mu^L, \\
                               \Phi_m(\infty) + \Phi\left(\frac{\bar{\tilde{k}}_\mu-\Lambda_L}{\sigma_L}\right) + \Phi\left(\frac{\bar{\tilde{k}}_\mu-\Lambda_H}{\sigma_H}\right) - 1 + \Phi_m\left(\frac{\Lambda_L-\bar{\tilde{k}}_\mu}{\sigma_L}\right) - \Phi_m\left(\frac{\bar{\tilde{k}}_\mu-\Lambda_H}{\sigma_H}\right) & \bar{\tilde{k}}_\mu^H<\bar{\tilde{k}}_\mu^L<\bar{\tilde{k}}_\mu,
                              \end{cases} 
\end{equation}
The pdf is found by differentiation to be:
\begin{eqnarray}
f(\bar{\tilde{k}}_\mu|\mu') = \begin{cases} 
                               \Phi_m(\infty)\delta(\bar{\tilde{k}}_\mu) + 
\frac{1}{2}\frac{1}{\sqrt{2\pi}}\frac{1}{\sqrt{\bar{\tilde{k}}_\mu}}\exp\left[-\frac{1}{2}\left(\sqrt{\bar{\tilde{k}}_\mu}-\Lambda_y\right)^2\right]   n_\mu\left(\Lambda_y - \sqrt{\bar{\tilde{k}}_\mu}\right) & \\ + \frac{1}{2}\frac{1}{\sqrt{2\pi}}\frac{1}{\sqrt{\bar{\tilde{k}}_\mu}}\exp\left[-\frac{1}{2}\left(\sqrt{\bar{\tilde{k}}_\mu}+\Lambda_y\right)^2\right]   n_\mu\left(\Lambda_y + \sqrt{\bar{\tilde{k}}_\mu}\right)
& \bar{\tilde{k}}_\mu<\bar{\tilde{k}}_\mu^L<\bar{\tilde{k}}_\mu^H, \\

                               \Phi_m(\infty)\delta(\bar{\tilde{k}}_\mu) + 
\frac{1}{2}\frac{1}{\sqrt{2\pi}}\frac{1}{\sqrt{\bar{\tilde{k}}_\mu}}\exp\left[-\frac{1}{2}\left(\sqrt{\bar{\tilde{k}}_\mu}+\Lambda_y\right)^2\right]   n_\mu(\Lambda_y+\sqrt{\bar{\tilde{k}}_\mu})   & \\ + 
\frac{1}{\sigma_L\sqrt{2\pi}}\exp\left[-\frac{1}{2}\frac{\left(\bar{\tilde{k}}_\mu-\Lambda_L\right)^2}{\sigma_L^2}\right]
n_\mu\left(\frac{\Lambda_L-\bar{\tilde{k}}_\mu}{\sigma_L}\right)
 & \bar{\tilde{k}}_\mu^L<\bar{\tilde{k}}_\mu<\bar{\tilde{k}}_\mu^H, \\

                               \Phi_m(\infty)\delta(\bar{\tilde{k}}_\mu) + \frac{1}{\sigma_L\sqrt{2\pi}}\exp\left[-\frac{1}{2}\frac{\left(\bar{\tilde{k}}_\mu-\Lambda_L\right)^2}{\sigma_L^2}\right]n_\mu\left(\frac{\Lambda_L-\bar{\tilde{k}}_\mu}{\sigma_L}\right) & \\ +
 \frac{1}{\sigma_H\sqrt{2\pi}}\exp\left[-\frac{1}{2}\frac{\left(\bar{\tilde{k}}_\mu-\Lambda_H\right)^2}{\sigma_H^2}\right]n_\mu\left(\frac{\bar{\tilde{k}}_\mu-\Lambda_H}{\sigma_H}\right) 
& \bar{\tilde{k}}_\mu^L<\bar{\tilde{k}}_\mu^H<\bar{\tilde{k}}_\mu \\ 

                               \Phi_m(\infty)\delta(\bar{\tilde{k}}_\mu) + 
\frac{1}{2}\frac{1}{\sqrt{2\pi}}\frac{1}{\sqrt{\bar{\tilde{k}}_\mu}}\exp\left[-\frac{1}{2}\left(\sqrt{\bar{\tilde{k}}_\mu}-\Lambda_y\right)^2\right]   n_\mu\left(\Lambda_y - \sqrt{\bar{\tilde{k}}_\mu}\right) & \\ + \frac{1}{2}\frac{1}{\sqrt{2\pi}}\frac{1}{\sqrt{\bar{\tilde{k}}_\mu}}\exp\left[-\frac{1}{2}\left(\sqrt{\bar{\tilde{k}}_\mu}+\Lambda_y\right)^2\right]   n_\mu\left(\Lambda_y + \sqrt{\bar{\tilde{k}}_\mu}\right)
& \bar{\tilde{k}}_\mu<\bar{\tilde{k}}_\mu^H<\bar{\tilde{k}}_\mu^L, \\

                               \Phi_m(\infty)\delta(\bar{\tilde{k}}_\mu) + \frac{1}{2}\frac{1}{\sqrt{2\pi}}\frac{1}{\sqrt{\bar{\tilde{k}}_\mu}}\exp\left[-\frac{1}{2}\left(\sqrt{\bar{\tilde{k}}_\mu}-\Lambda_y\right)^2\right] n_\mu\left(\Lambda_y - \sqrt{\bar{\tilde{k}}_\mu}\right) & \\ +
\frac{1}{\sigma_H\sqrt{2\pi}}\exp\left[-\frac{1}{2}\frac{\left(\bar{\tilde{k}}_\mu-\Lambda_H\right)^2}{\sigma_H^2}\right] n_\mu\left(\frac{\bar{\tilde{k}}_\mu-\Lambda_H}{\sigma_H}\right)
& \bar{\tilde{k}}_\mu^H<\bar{\tilde{k}}_\mu<\bar{\tilde{k}}_\mu^L, \\

                               \Phi_m(\infty)\delta(\bar{\tilde{k}}_\mu) + \frac{1}{\sigma_L\sqrt{2\pi}}\exp\left[-\frac{1}{2}\frac{\left(\bar{\tilde{k}}_\mu-\Lambda_L\right)^2}{\sigma_L^2}\right]n_\mu\left(\frac{\Lambda_L-\bar{\tilde{k}}_\mu}{\sigma_L}\right) & \\ +
 \frac{1}{\sigma_H\sqrt{2\pi}}\exp\left[-\frac{1}{2}\frac{\left(\bar{\tilde{k}}_\mu-\Lambda_H\right)^2}{\sigma_H^2}\right]n_\mu\left(\frac{\bar{\tilde{k}}_\mu-\Lambda_H}{\sigma_H}\right) 
& \bar{\tilde{k}}_\mu^H<\bar{\tilde{k}}_\mu^L<\bar{\tilde{k}}_\mu,
                              \end{cases}
\end{eqnarray}
where 
\begin{equation}
 n_\mu(x) = 1 - m_\mu(x\sigma + \mu').
\end{equation}
The formulae presented in equations~\ref{cumultwosideddist} and~\ref{twosideddist} are the same as the above formulae, expressed in a more compact notation using the discrete Heaviside function.

\section{Conclusion}
The asymptotic distribution for a general likelihood-based test statistic is derived, which covers all previously described asymptotic distributions as special cases of the general formula presented here. 

\section{Acknowledgements}
The author expresses his thanks to C. Lester for reviewing this paper and his useful suggestions.

\end{document}